# Quasicrystalline Nanocrystal Superlattice with Partial Matching Rules

Xingchen Ye[1,†], Jun Chen[1,2,†], M. Eric Irrgang[3,4,†], Michael Engel[4,5,†], Angang Dong[6], Sharon C. Glotzer[3,4,5,*], Christopher B. Murray[1,2,*]

[1]Department of Chemistry, [2]Department of Materials Science and Engineering, University of Pennsylvania, Philadelphia, Pennsylvania 19104, USA.
[3]Department of Materials Science and Engineering, [4]Biointerfaces Institute, [5]Department of Chemical Engineering, University of Michigan, Ann Arbor, Michigan 48109, USA.
[6]Department of Chemistry and Collaborative Innovation Center of Chemistry for Energy Materials, Fudan University, Shanghai 200433, China.

†These authors contributed equally.
*Corresponding authors: sglotzer@umich.edu, cbmurray@sas.upenn.edu

**Expanding the library of self-assembled superstructures provides insight into the behavior of atomic crystals and supports the development of materials with mesoscale order[1,2]. Here we build upon recent findings of soft matter quasicrystals[3–6] and report a quasicrystalline binary nanocrystal superlattice that exhibits correlations in the form of partial matching rules reducing tiling disorder. We determine a three-dimensional structure model through electron tomography[7,8] and direct imaging of surface topography. The 12-fold rotational symmetry of the quasicrystal is broken in sub-layers, forming a random tiling of rectangles, large triangles, and small triangles with 6-fold symmetry. We analyze the geometry of the experimental tiling and discuss factors relevant for the stabilization of the quasicrystal. Our joint experimental-computational study demonstrates the power of nanocrystal superlattice engineering and further narrows the gap between the richness of crystal structures found with atoms and in soft matter assemblies.**

Monodisperse colloidal nanocrystals (NCs), often regarded as artificial atoms, self-assemble into a rich variety of simple or exotic superlattices[9] including quasicrystals[6,10]. These quasicrystals are a subset of the more general class of soft matter quasicrystals, which correspond to two-dimensional tilings when projected along the periodic direction. The tile edge length spans two orders of magnitude from pentameric molecules[3] (10-fold symmetry, tile edge length 0.82 nm) via dendrimeric macromolecules[4] (12-fold, 8.2 nm) to ABC star polymers[5] (12-fold, 80 nm). Understanding stability and designing new quasicrystals requires targeting the right local order found in tiles and an appropriate arrangement of the tiles, which can be controlled systematically by two competing length scales[11] or appropriate bond angles[12]. Two extreme scenarios are distinguished: matching rules enforce quasiperiodicity by penalizing incorrect tile attachments[13], and random tilings embrace tile reshuffling and predict the emergence of a quasicrystal due to entropy maximization[14]. In both cases, growth can proceed by a rapid stochastic process followed by a slower error-and-repair process[15]. The structural quality depends on the strength of tile interactions and the time available for the repair process.

Matching rules control the growth process of periodic and aperiodic tilings[16] from nanocrystal superlattices to DNA origami[17]. They are particularly important for dodecagonal (12-fold) quasicrystals (DDQCs), which typically feature an aperiodic tiling of squares and equilateral triangles[18]. Squares and triangles easily tile the plane and form ordered or disordered tilings[19,20]. It has been suggested that to achieve well-ordered square-triangle tilings, certain tile contacts or vertex configurations should be biased by introducing additional correlations[21,22]. We call this process 'partial matching rules' if the correlations improve quasiperiodic ordering but are too weak to enforce perfect order. Here, we report a type of DDQC in self-assembled binary nanocrystal superlattices (BNSLs) in which the decoration of the square and triangle tiles naturally gives rise to partial matching rules. We analyze the quality of the DDQC by performing a phason displacement field analysis, investigate the origin of stability, and formulate a model that reproduces its growth via molecular dynamics simulation.

The DDQCs formed upon slow drying of binary solutions of NCs (Supplementary Figs. 1-2) on top of an immiscible liquid subphase[23] as thin-film (about 100 nm thick) structures with lateral dimensions up to the centimeter scale. Fig. 1a shows a large-area transmission electron microscopy (TEM) image of BNSLs self-assembled from 6.8 nm $CoFe_2O_4$ and 12.0 nm $Fe_3O_4$ NCs (Supplementary Fig. 3). The BNSLs possess long-range order, yet no translational periodicity. The small-angle electron diffraction (SAED) pattern has diffraction spots with 12-fold rotational symmetry (Fig. 1b), the characteristic signature of DDQCs. The spots are sharp and higher-order reflections are present, which indicates that the structural perfection persists over length scales exceeding prior observations of DDQCs in BNSLs[6,10]. Weak peaks are suppressed due to the random tiling character of the DDQC. The high-magnification TEM image shows that the DDQC is an arrangement of square and triangle tiles (Fig. 1c). We also realized the same DDQC in BNSLs composed of FePt and $Fe_3O_4$ NCs (Fig. 1d,e and Supplementary Fig. 4), $Fe_3O_4$ and $Fe_3O_4$ NCs (Supplementary Fig. 5), and 5.8 nm Au and 9.7 nm $Fe_3O_4$ NCs (Supplementary Fig. 6). In all realizations, the NC size ratio lies in the range 0.60-0.66. Here the NC size is computed as the sum of the inorganic core diameter plus the ligand shell thickness. The observation of the DDQC in four independent systems demonstrates that its formation is robust and does not depend significantly on details of the constituent NCs. In line with previous studies[6,10], the present DDQC can coexist with its σ-phase approximant[24] (Fig. 1f). Both phases intergrow in the $CoFe_2O_4$-$Fe_3O_4$ system (Fig. 1g and Supplementary Figs. 7-8) and the FePt-$Fe_3O_4$ system (Supplementary Figs. 9-10). The σ-phase corresponds to a periodic $3^2.4.3.4$ Archimedean square-triangle tiling. It is locally similar to a DDQC but differs in its global ordering. Other periodic BNSLs of $NaZn_{13}$ type and $CaCu_5$ type coexist with the DDQC and the σ-phase but occur less frequently (Supplementary Figs. 11-12). Such coexistence has been proposed and experimentally observed in complex supramolecular assemblies[4,25].

Although the square-triangle tiling is consistent with past reports of DDQCs in BNSLs, the tile decoration is distinct from known BNSL unit cells. The presence of overlapping NC layers in TEM images suggests a complex three-dimensional structure that cannot be resolved through visual inspection of projections alone. Instead we employ electron tomography[7,8]. The tomographic reconstruction of the $CoFe_2O_4$-$Fe_3O_4$ DDQC clearly resolves the positions and type (large or small) of individual NCs within individual layers (Fig. 2a-e (top), Supplementary Figs. 13-14 and Supplementary Movies 1-7). For reasons discussed below, we distinguish mirror (M, M′), puckered (P, P′), and tiling (T) layers. Vertices defining the square-triangle tiling correspond to stacks of large NCs (Fig. 2a-e (middle)). The symmetry of each layer is easily determined from the Fourier transforms of the tomography slices (Fig. 2a-e (bottom)). Interestingly, the 12-fold symmetry of the overall structure is broken into 6-fold symmetry in all layers except the T layer. The orientation of the 6-fold symmetric layers rotates by 30 degrees between the non-primed (P, M) and primed (P′, M′) layers, confirming the existence of a screw symmetry. Such a symmetry breaking has not yet been observed in soft matter quasicrystals.

Closer inspection of the tomographic reconstruction of the M layer reveals an aperiodic tiling consisting of rectangles and equilateral triangles of two different edge lengths with the edge lengths of the rectangles equal to those of small and large triangles (Fig. 2a). The rectangle tile is composed of four large NCs as vertices and one small NC at the center. The large triangle tile is made up of three large NCs and one small NC at the center. The center NC is absent for small triangle tiles. High-resolution scanning electron microscopy (SEM) imaging of DDQC surface topography reveals the same rectangle-large triangle-small triangle tiling in both $CoFe_2O_4$-$Fe_3O_4$ and $FePt$-$Fe_3O_4$ DDQCs and the σ-phase (Fig. 2f-l). We observe three surface terminations from different domains of DDQCs, which correlate well with tomographic reconstruction (Fig. 2i-k and Supplementary Figs. 15-17).

Based on electron tomography and SEM results, we construct a structure model of the DDQC (Fig. 3a and Supplementary HTML visualization). It is now apparent that mirror layers (M, M′) in fact coincide with mirror planes, puckered layers (P, P′) are slightly puckered, and the tiling layer (T) contains NCs sitting at the vertices of the square-triangle tiling. Over the range of one periodicity, square tiles contain 2 large and 16 small NCs (composition $AB_8$). Triangle tiles contain 1 large and 7 small NCs ($AB_7$). The σ-phase unit cell then contains 8 large and 60 small NCs ($AB_{7.5}$). For a full stacking sequence M, P, T, P′, M′, P′, T, P, M of height $z = 1$ and M′ at $z = 1/2$, the P/P′ and T layers are not evenly spaced vertically in the cell but lie at approximately $z = 2/12$, $3/12$, $4/12$ and $z = 8/12$, $9/12$, $10/12$. This is evident both from packing considerations and by examination of the tomography results. The large NCs occurring in M and M′ layers form staggered columns. Lines connecting large NCs to neighbors in the same layer give rise to edges of two lengths that bound small and large triangles as well as rectangles. Long and short edges lie along two hexagonal subsets of the twelve long diagonals of a dodecagon. The two subsets are rotated relative to one another by 30 degrees and alternate between M and M′ layers, indicative of a $12_6$ screw symmetry. The result is a tiling of square tiles consisting of layers of rectangles with alternating orientations and triangle tiles containing layers of alternating large and small triangles. Based on the structure model, we postulate a point group 12/*mmm* and five-dimensional space group P12$_6$/*mcm*. This space group is distinct from known BNSL DDQCs[6] but has recently been reported in atomic systems[26].

Although the square and triangle decorations in Fig. 3a are developed geometrically with maximal symmetry, some symmetry is lost with the internal strain necessary for different neighbor relationships. For instance, in the projection of the σ-phase, alternating layers of large NCs lie along the shared edge of two triangles rather than shifting radially from the tile centers (Fig. 3b). Depending on the environment, one or more columns of large NCs may be pinned to a tile vertex, as in a hexagon consisting of six triangle tiles (Fig. 3c). Internal strain can also be relaxed through defects, e.g. the replacement of a large NC by a small one, which was observed occasionally in the electron tomography data. The distribution of the two edge types (long and short) in the M and M′ layers is governed by two rules: (1) triangle edges are of the same type; (2) square edges alternate in type. As can be shown, these rules are applicable to any square-triangle tiling. The result is a grouping of edges into short and long based on their orientation alone. This explains the breaking of the 12-fold symmetry into the 6-fold symmetry observed in individual M and P layers. 12-fold symmetry is observed only in the T layer or when averaging all layers by projection. The unusual decoration of the square and triangle tiles affects the tiling geometry. Joining two triangles or two squares along an edge or more than two triangles without the separation of a square around a vertex introduces internal stress, and thus penalties in the form of partial matching rules. As a consequence, the randomness of the tiling is decreased.

We quantify the amount of randomness by analyzing the phason displacement field of the quasiperiodic tiling. Such an analysis was performed previously for an icosahedral quasicrystal formed in computer simulation[27] and for DDQCs in a Mn-Cr-Ni-Si alloy[26] but not yet for soft matter quasicrystals. We focus on CoFe$_2$O$_4$-Fe$_3$O$_4$ samples for which automated image processing is reasonably effective to identify the tiling. Image processing extracted as many as 15,000 vertices from the lowest magnification TEM images (Fig. 3d). The identified vertices are lifted into a four-dimensional configuration space by assigning each vertex a four-dimensional lattice point, which is then projected onto a two-dimensional parallel space, $x_j^{\|}$, describing the phonon strain-corrected position of the vertex, as well as onto the two-dimensional perpendicular space, $x_j^\perp$, quantifying the amount of phason displacement (Supplementary Figs. 18-23). The lifting procedure works well in our samples except for the presence of occasional dislocations (Fig. 3e and Supplementary Figs. 24-25). In agreement with prior works[28], the dislocations are predominantly of one type, though the Burgers vector we observe most frequently, $\boldsymbol{b} = (0,-1,2,-1)$, is different.

The phason displacement analysis correlates the distances of two vertices in parallel space, $r_{jk}^{\|} = |x_j^{\|} - x_k^{\|}|$, and perpendicular space, $r_{jk}^\perp = |x_j^\perp - x_k^\perp|$. We analyze three samples and compare them with an ideal square-triangle DDQC constructed via an inflation rule (Fig. 3f and Supplementary Fig. 26). The comparison shows that the phason displacement in our samples is clearly distinct from the linear slope of the σ-phase, $r^\perp \propto r^{\|}$, and instead follows the behavior expected for two-dimensional random tilings[14] by deviating only logarithmically from the ideal tiling, $r^\perp = K^{-1}\ln\left(r^{\|}/a\right) + C$ with tile edge length $a$ and constants $C$ and $K$.

To understand the stabilization of the DDQC, we consider its feasibility due solely to geometric effects as a binary packing of hard particles. We model NC shapes as spheropolyhedra interpolating between a sphere and a truncated octahedron. Particles of that shape fit together well in the tiling decoration of the DDQC (Fig. 4a). Packing considerations had been successful in describing the ordering of binary sphere[29] and rod-sphere[30] mixtures. Here, however, the phase-separated state always has a significantly higher packing density (Fig. 4b and Supplementary Figs. 27-28). Our results suggest that although a good circumradius ratio lies between 0.59 and 0.63, which is close to the NC size ratio used in experiment, entropy alone is not sufficient to stabilize the DDQC at any pressure.

Next, we undertake molecular dynamics simulations in an attempt to stabilize the DDQC using pair interactions. We pursue an isotropic binary particle model at the size ratio suggested by the packing density analysis and by experiments. To promote mixing, we combine excluded volume repulsions with an A-B attraction via a Morse potential. A fluid is seeded with several unit cells of the σ-phase. We observe growth of the seed for a narrow range of parameters (Fig. 4c-e and Supplementary Movie 8). After initial rapid growth along the 12-fold axis, growth is significantly slower perpendicular to this axis. These results confirm the soundness of the structure model in Fig. 3a. Furthermore, our simulations suggest that particle size ratio and the width of the attraction well are crucial factors in stabilizing the DDQC over competing phases.

In summary, we have discovered a quasicrystalline BNSL with structural complexity exceeding previously reported nanocrystal and colloidal assemblies. The NC decoration of the tiles has no analogue in known periodic or aperiodic BNSLs. We believe that achieving partial matching rules will be key when searching for new complex nano- and mesoscale assemblies.

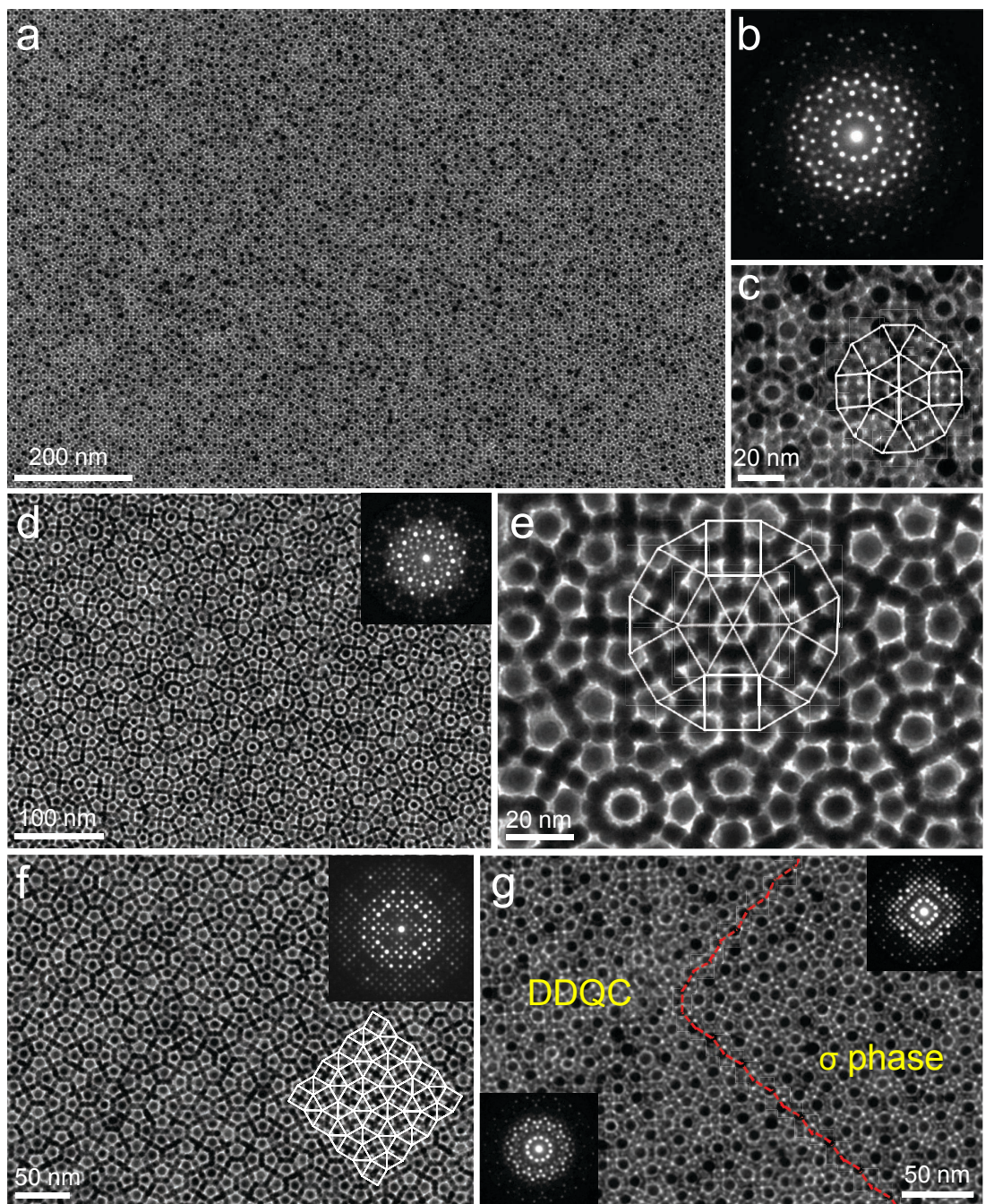

**Figure 1. Self-assembled binary nanocrystal superlattices with quasiperiodic and periodic order.** (a) Low-magnification TEM image, (b) SAED pattern and (c) high-magnification TEM image of a dodecagonal quasicrystal self-assembled from 6.8 nm $CoFe_2O_4$ and 12.0 nm $Fe_3O_4$ nanocrystals. (d) Low-magnification TEM image and SAED pattern (inset) and (e) high-magnification TEM image of the quasicrystal self-assembled from 6.2 nm FePt and 11.5 nm $Fe_3O_4$ nanocrystals. (f) TEM image and SAED pattern (inset) of the competing periodic σ-phase in a FePt-$CoFe_2O_4$ superlattice. The $3^2.4.3.4$ tiling of the σ-phase is shown as an overlay. (g) TEM image and SAED patterns (insets) show the coexistence of the quasicrystal and the σ-phase in a $CoFe_2O_4$-$Fe_3O_4$ superlattice.

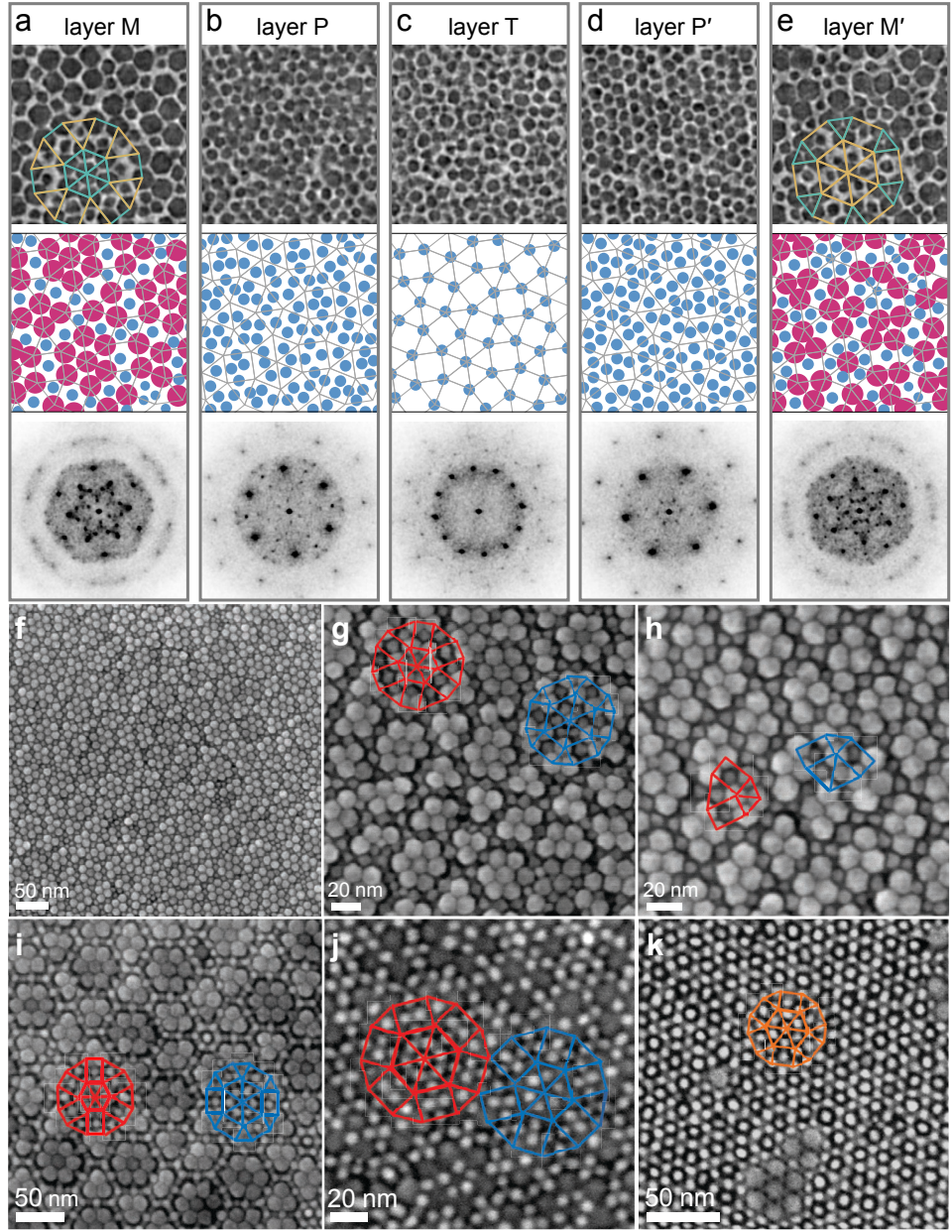

**Figure 2. Tomographic reconstruction and SEM imaging reveal the three-dimensional structure of the superlattices.** (a-e) Horizontal reconstruction slices show the nanocrystal arrangement within the layers M, P, T, P′, M′ at different heights via tomographic slices (top row), the derived decoration of the tiling (middle), and the Fourier transform of the tomographic slices (bottom). Except for layer T, all layers exhibit 6-fold rotational symmetry. The slices are taken from the tomographic reconstruction data shown in Supplementary Movie 1. SEM image of (f,g) the quasicrystal and (h) the σ-phase in $CoFe_2O_4$-$Fe_3O_4$ superlattices. (i-k) SEM image of the quasicrystal in $FePt$-$Fe_3O_4$ superlattices. We observe three distinct surface terminations: M layer (i), M and P layer combined (j), P and T layer combined (k). The red and blue overlays highlight two distinct tiling motifs that are related by swapping large and small triangles while rotating the motif by 30 degrees.

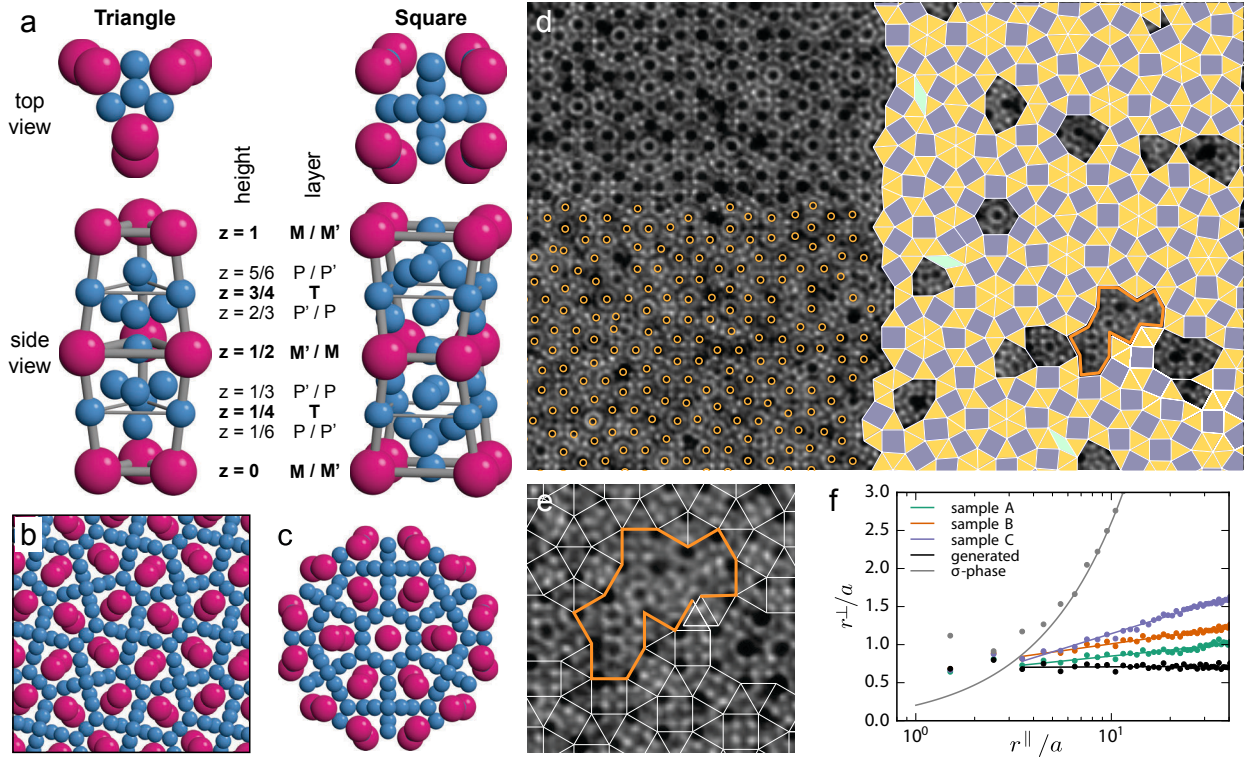

**Figure 3. Three-dimensional structure model of the quasicrystal and geometric tiling analysis.** (a) The quasicrystal can be described as a decorated tiling of equilateral triangle tiles and square tiles. Nanocrystals align on flat mirror layers (M, M′), flat layers that contain small particles at tile vertices (T), and puckered layers (P, P′). The tile decoration with large and small particles adjusts to the local tiling environment as shown here for (b) the σ-phase and (c) a dodecagon ring. (d) Columns of large particles are identified with image processing (bottom left) and mapped to vertices of a square-triangle tiling (right). The mapping is hindered by the occasional presence of a dislocation (highlighted in orange). (e) The Burgers vector is identified in idealized coordinates by performing a Burgers circuit around the dislocation core using a four-dimensional basis. (f) We analyze the tiling for the presence of phason displacement by correlating average separation in parallel space $r^{\parallel}$ with separation in perpendicular space $r^{\perp}$ in units of tile edge length $a$. Phason fluctuations vary in three analyzed samples (denoted A-C) and are lowest in sample A.

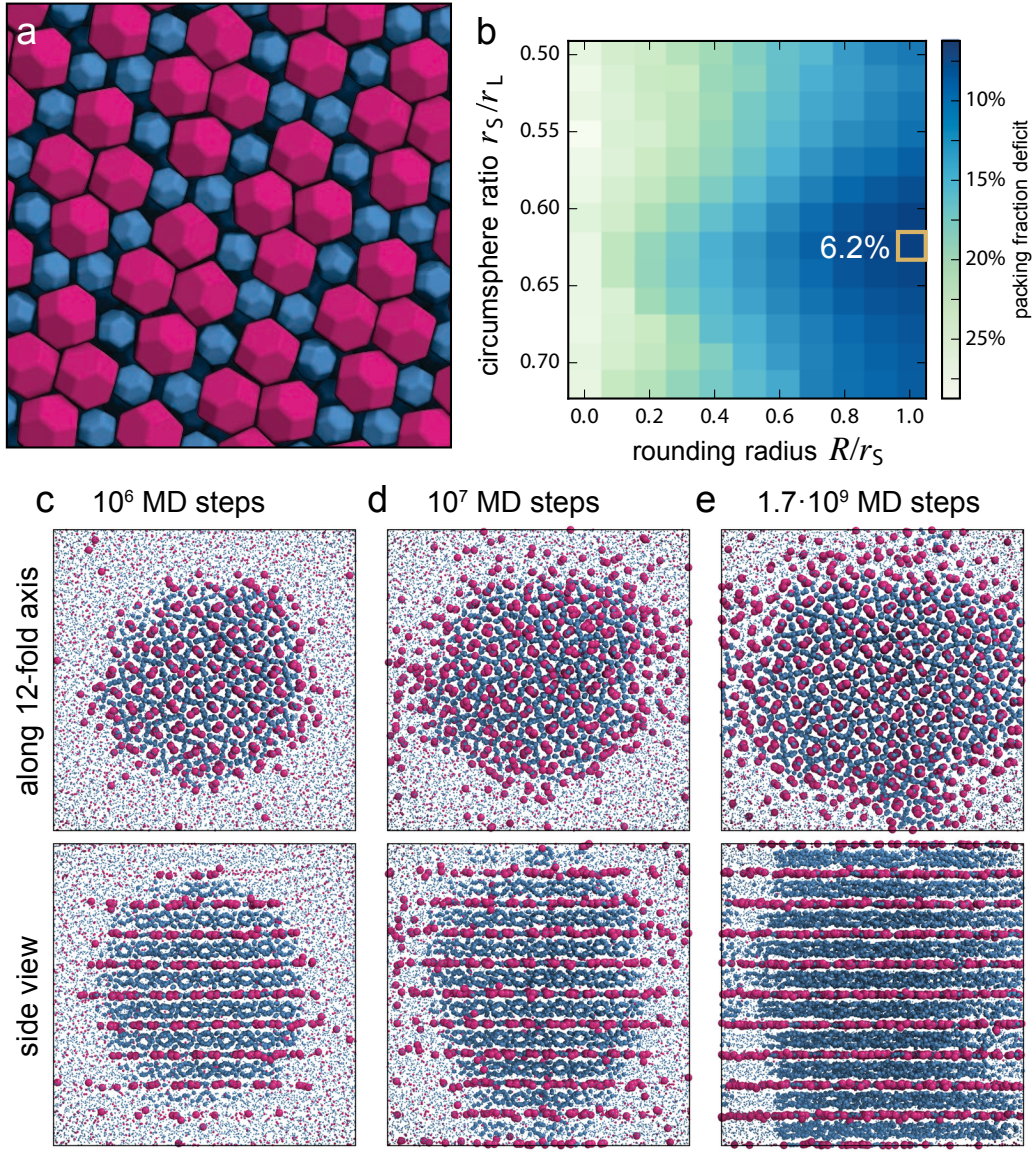

**Figure 4. Stabilization and growth of the quasicrystal.** (a) Spherotruncated octahedral particles found in the $CoFe_2O_4$-$Fe_3O_4$ system are positioned on the decoration model for the σ-phase and compressed numerically assuming excluded volume interactions only. The particles show well defined but not unique orientations. (b) Particle size ratio (measured as the ratio of circumscribing spheres) and rounding radius were varied in an attempt to discover good particle geometries that entropically favor the formation of the quasicrystalline superlattice. Packing fraction deficit is defined as the difference of the packing fraction of the densest lattice packing of individual particles (phase separated state) and the packing fraction of the densest packing of the σ-phase in the limit of infinite pressure (binary state). The optimization suggests an ideal size ratio between 0.59 and 0.63. (c-e) Observation of quasicrystal growth from a seed in a molecular dynamics (MD) simulation of isotropic particles with A-B attraction at particle size ratio 0.56. Particles are identified as fluid if they move rapidly relative to their neighbors and then shown as small dots to distinguish them from the particles in the growing solid.

**Acknowledgements**

X.Y. and C.B.M. were supported by the U.S. Department of Energy Office of Basic Energy Sciences, Division of Materials Science and Engineering under Award No. DE-SC0002158. J.C. and C.B.M. received support from NSF MRSEC under Award No. DMR-1120901. M.E.I., M.E., and S.C.G. were supported by the U.S. Army Research Office under Award No. W911NF-10-1-0518 and by the Assistant Secretary of Defense for Research and Engineering, U.S. Department of Defense under Award No. N00244-09-1-0062. Any opinions, findings, and conclusions or recommendations expressed in this publication are those of the authors and do not necessarily reflect the views of the DOD/ASD(R&E). S.C.G was partially supported by a Simons Investigator award from the Simons Foundation. We thank Ben Schultz for providing image analysis code. This work used the Extreme Science and Engineering Discovery Environment (XSEDE), which is supported by NSF grant number ACI-1053575, XSEDE award DMR 140129. Additional computational resources and services were supported by Advanced Research Computing at the University of Michigan, Ann Arbor.

**Author contributions**

X.Y., J.C., and C.B.M. conceived and designed the experiments. X.Y. and J.C. carried out nanocrystal syntheses and self-assembly, and carried out TEM imaging and electron tomography reconstruction. X.Y., J.C., and A.D. performed SEM imaging. M.E.I., M.E., and SCG planned and discussed theoretical analysis and computer simulations. M.E.I. performed analysis and simulations. M.E., S.C.G., and C.B.M. supervised the project. All authors discussed the results and commented on the manuscript.

**Competing financial interests**

The authors declare no competing financial interests.

**Methods**

**Nanocrystal synthesis and self-assembly.** $Fe_3O_4$ NCs[31], $CoFe_2O_4$ NCs[32], FePt NCs[33], and Au NCs[34] were synthesized according to literature methods. All NC syntheses were carried out under nitrogen atmosphere using standard Schlenk line techniques. The as-synthesized NCs were precipitated with ethanol, isolated using centrifugation, and redispersed in hexane. BNSLs were formed by co-crystallization of a binary mixture of NCs at room temperature on top of the immiscible liquid subphase diethylene glycol (DEG). In a typical process, a hexane solution (15 μL) containing two types of NCs was drop-cast onto the surface of DEG in a Teflon well (1.5 x 1.5 x 1.5 $cm^3$). The well was then covered with a glass slide to slow down solvent evaporation. After 30 min, the BNSL film was transferred onto a carbon-coated Cu TEM grid (300-mesh), which was further dried under vacuum to remove residual DEG.

**Electron microscopy.** TEM images, SAED and wide-angle electron diffraction (WAED) patterns were taken on a JEOL JEM-1400 transmission electron microscope equipped with a SC 1000 ORIUS CCD camera operating at 120 kV. HRSEM images were obtained on a JEOL 7500F SEM operating at 5kV.

**Electron tomography.** The projection images for electron tomography were acquired on a JEOL JEM-1400 operating in bright-field TEM mode at 120 kV. A dilute aqueous solution of citrate-stabilized Au NCs (~15 nm in diameter) was drop-cast onto BNSLs prior to image acquisition. Images were recorded over an angular range of ±60° at 1° increments. The tilt series was aligned using either citrate-stabilized Au NCs or isolated NCs from the BNSL sample as fiducial markers and reconstructed using the eTomo software of the IMOD tomography package[35,36].

**Extraction of tile vertices.** Dark circular spots of relatively uniform size in TEM images mark tile vertices that were identified using image processing software[37,38]. The centers of the spots were accentuated by convolution with a disk of the same size. A threshold separated the peaks, then a watershed cut built clusters of pixels associated with each local maximum. Automatic tile vertex detection works well for $CoFe_2O_4$-$Fe_3O_4$ systems, where the vertices have higher contrast in TEM, but not for samples in which the small particles are FePt NCs.

**Lifting of tile vertices.** DDQCs were described as tilings of squares, triangles, and thin 30 degree rhombi (low occurrence). We assigned each vertex $x_i$ quasi-lattice coordinates $x_i^{(4)}$ in the form of integer linear combinations of the basis vectors $e_j = P^{\parallel} e_j^{(4)} = (\cos(j\pi/6), \sin(j\pi/6))a, j =$ 1,2,3,4 for higher-dimensional embedding[39]. Tile vertices were identified by breadth-first network exploration following connected edges using a tolerance $0.1a$. Stretch, shear, and rotation due to the imperfect perpendicularity of the sample to the electron beam were corrected by linear regression. We identified as square, triangle, and rhomb tiles any closed loop of three or four points. An ideal square-triangle DDQC was constructed via inflation rule[40].

**Densest packing.** Simulation boxes of the σ-phase unit cell were repeatedly compressed to high pressure and relaxed using Monte Carlo simulation following the procedure in our past work[41]. Overlap checks consider particles as the Minkowski sum of a sphere and an Archimedean truncated octahedron[42,43].

**Molecular dynamics simulations.** Molecular dynamics simulations were performed using the open-source HOOMD-blue simulation package[44,45]. We assume point particles with an attractive Morse potential to represent an orientationally averaged interparticle interaction. The code may be downloaded from the online repository at http://glotzerlab.engin.umich.edu/hoomd-blue/. Simulations were run on the XSEDE computing infrastructure[46].

**Additional References**